\documentstyle[multicol,prl,aps]{revtex}

\begin{document}
\draft
\title{Classical properties of low-dimensional conductors:\\
Giant capacitance and non-Ohmic potential drop}
\author{Boris Korenblum$^1$ and Emmanuel I.~Rashba$^2$\cite{Rashba*}}
\address{$^1$Department of Mathematics and Statistics, SUNY at Albany, Albany, NY 12222\\
$^2$Department of Physics, SUNY at Buffalo, Buffalo, NY 14260}
\date{April 17, 2002}
\maketitle
\begin{abstract} 
Electrical field arising around an inhomogeneous conductor when an electrical current passes through it is not screened, as distinct from 3D conductors, in low-dimensional conductors. As a result, the electrical field depends on the global distribution of the conductivity $\sigma(x)$ rather than on the local value of it, inhomogeneities of  $\sigma(x)$ produce giant capacitances $C(\omega)$ that show frequency dependence at relatively low $\omega$, and electrical fields develop in vast regions around the inhomogeneities of $\sigma(x)$. A theory of these phenomena  is presented for 2D conductors. 
               
\end{abstract}
\pacs{PACS numbers: 73.63.-b,73.50.-h,73.23.-b}

\begin{center}
{\it Editorially approved, to appear in Phys. Rev. Lett.}
\end{center}
\begin{multicols}{2}

Quantum confinement of electrons in low-dimensional (LD) conductors results in dramatic suppression of the electrical screening that changes from the exponential in 3D to the Stern power-law screening in 2D\cite{Stern}. It manifests itself, e.g., in the softening of plasma modes \cite{MV}. Screening phenomena should also strongly influence the passage of currents through the circuits including inhomogeneous LD conductors. It is well understood now that the applicability criteria of the four- and two-terminal Landauer-B\"{u}ttiker equations\cite{LB} are controlled by the electrical screening \cite{TLKK}. The latter is also of importance for calculating spin-injection transients \cite{R02} and for some different problems of  spintronic devices \cite{spintr}. However, calculating the effect of screening on the potential distribution around a long resistive LD conductor with a potential moderately growing at infinity needs the techniques that differ from the standard techniques used, e.g., for plasmons. Meantime, in addition to the horizontal transport in narrow quantum wells (remarkably, few-nanometer thick strips\cite{fewnano} have been created) and in single-wall nanotubes \cite{fewnano,nanotubes}, electrical transport through a chain of individual gold atoms \cite{Au} and even through individual molecules \cite{SMB} has been reported. Under these conditions developing a regular procedure for calculating the effect of screening on the parameters of extended LD conductors becomes an important issue.

In regular 3D conductors the screening length $l_{3D}$ is small compared to all geometrical dimensions. Therefore, the inhomogeneous electrical field produced in vacuum (or an insulator) around the conductor by the potential drop inside it is completely screened in the narrow surface layer about $l_{3D}$ and does not penetrate into the bulk of the conductor. As a result, there exists the usual {\it local Ohmic relation} between the field ${\bf E}({\bf r})$ and the current density ${\bf j}({\bf r})$, ${\bf E}({\bf r})={\bf j}({\bf r})/\sigma({\bf r})$, where $\sigma({\bf r})$ is the conductivity (the mean free path $l_{\rm m.f.p.}$ is assumed to be small enough). This is not the case for LD conductors with long-range screening. Therefore, the relation between ${\bf j}({\bf r})$ and ${\bf E}({\bf r}^{\prime})$ becomes nonlocal.

We show that for LD conductors:

(i) the potential drop is non-Ohmic because the field ${\bf E}({\bf r})$ is controlled by the global distribution of $\sigma({\bf r})$ rather than by its local value,

(ii) there is no macroscopic limit for the potential drop per unit length for a long LD conductor, and

(iii) inhomogeneous ``Ohmic" LD conductors acquire a giant capacitance that shows dispersion at low frequencies and can be measured in ac experiments.

To make the essence of the phenomena most clear we choose the simplest model of a conductor in the diffusive regime when a local constitutive equation ${\bf j}({\bf r})=\sigma({\bf r})\nabla \zeta({\bf r})$ relating ${\bf j}({\bf r})$ to the gradient of the electrochemical potential $\zeta({\bf r})$ holds. The 2D conductor resides in the $xz$-plane, its 2D conductivity $\sigma({\bf r})=\sigma(x)$, and the current ${\bf j}({\bf r})\| {\hat x}$. The electrical potential $\varphi(x,y)$ satisfies the Laplace equation $\Delta\varphi(x,y)=0$  for $y\neq 0$ and is continuous at the conductor boundary. The boundary condition follows from the equation
 \begin{equation}
n(x)=e\rho[\zeta(x)+\varphi(x)],~~\varphi(x)\equiv\varphi(x,y=0),
\label{eq3}
\end{equation}
that is fulfilled inside the LD conductor in the linear approximation in the current $j(x)$. Eq.~(\ref{eq3}) relates $\zeta(x)$ and $\varphi(x)$ to the nonequilibrium 2D electron concentration in the conductor, $n(x)$, and the 2D density-of-states in it, $\rho$, that we assume to be $x$-independent. For the upper half-plane, $y>0$, the boundary condition for the Laplace problem, according to the Gauss theorem, reads as
\begin{equation}
l\varphi^{\prime}_y(x)-\varphi(x)=\zeta(x),~~
\varphi^{\prime}_y(x)\equiv\partial_y\varphi(x,y)|_{y\rightarrow 0+}~,
\label{eq4}
\end{equation}
where $l=\varepsilon/2\pi e^2\rho$ is the 2D screening length, $\varepsilon$ being the dielectric constant of the insulator surrounding the 2D conductor. The ratio $l/l_{3D}\sim\sqrt{r_Bk_F}/k_Fw$, where $r_B$ is the Bohr radius, $k_F$ is the Fermi momentum, and $w$ is the thickness of the 2D layer (in the $y$-direction). In a quantum conductor this ratio is large, $l/l_{3D}\gg 1$, because $r_Bk_F\agt 1$ for a metallic conductor, and $k_Fw\ll 1$ under the conditions of a strong 2D confinement.

Therefore, the Laplace equation should be solved with the mixed boundary condition of Eq.~(\ref{eq4}). Let us introduce an auxiliary function
  \begin{equation}
\psi(x,y)={y\over\pi}\int_{-\infty}^{\infty}{{\zeta(t)~dt}\over{(t-x)^2+y^2}}=
{1\over\pi}{\rm Im}\int_{-\infty}^{\infty}{{\zeta(t)}\over{t-z}}dt,
\label{eq5}
\end{equation}
$z=x+iy$, that is harmonic for $y>0$ and satisfies the Dirichlet boundary condition $\psi(x,0)=\zeta(x)$ provided the integral exists. Such an approach is advantageous compared to the standard Fourier transformation technique because it works for functions $\zeta(x)$ moderately increasing at infinity. For $\zeta(x)\propto \vert x\vert^\alpha{\rm sign} \{x\}$ the integral in Eq.~(\ref{eq5}) converges for $\alpha<1$ and diverges as a power of the lead length for $\alpha>1$. The case of Ohmic leads, $\alpha=1$, is marginal and results in the logarithmic singularities discussed below. Then the function 
 \begin{eqnarray}
\varphi(x,y)&=&-e^{y/{\it l}}\int_y^\infty\psi(x,t)e^{-t/{\it l}}~{{dt}\over l} \nonumber\\
&=&-\int_0^\infty\psi(x,y+t)e^{-t/{\it l}}~{{dt}\over{\it l}}
\label{eq6}
\end{eqnarray}
is also harmonic in the upper half-plane, $y>0$, and satisfies the mixed boundary condition of Eq.~(\ref{eq4}). Because $\zeta^{\prime}(x)=j/\sigma(x)$ for an one-dimensional flow, the function $\zeta(x)$ is known and Eq.~(6) provides an explicit solution for the potential $\varphi(x,y)$. 

To find out the basic regularities of the potential distribution inside LD conductors (and around them) let us consider several model systems. For a resistive contact at $x=0$ in a perfect conductor, $1/\sigma(x)=\delta(x)/\Sigma$, the electrochemical potential equals $\zeta(x)=(j/\Sigma)\Theta(x)$, where $\Sigma^{-1}$ is the contact resistance and $\Theta(x)$ is the Heaviside function. It follows from Eqs.~(\ref{eq5}) and (\ref{eq6}) that
  \begin{eqnarray}
\psi(x,y)&=&(j/\pi\Sigma)~{\rm arccot}(-x/y),\nonumber\\
\varphi(x)&=&-{j\over\Sigma}\Theta(x)+{{jx}\over{\pi l\Sigma}}
\int_0^\infty{{e^{-\tau}d\tau}\over{\tau^2+(x/l)^2}}~.
\label{eq7}
\end{eqnarray}
As distinct from $\zeta(x)$, the potential $\varphi(x)$ is continuous near $x=0$ but acquires a singular contribution $(jx/\pi l\Sigma)\ln(|x|/l)$ \cite{GrRy}. Therefore, the electrical field diverges near $x=0$ as $E_x(x)\approx (j/\pi l\Sigma)\ln(|x|/l)$ inside the conductor and in the vicinity of it. For $|x|>>l$ the potential $\varphi(x)$ approaches its asymptotical values slowly, as $\varphi(x)\approx-\zeta(x)+jl/\pi\Sigma x$. The total drop in $\varphi(x)$ is equal to the drop in $\zeta(x)$. However, while the drop in  $\zeta(x)$ occurs abruptly at $x=0$, the potential $\varphi(x)$ changes gradually at the scale of $l$.

With the $\varphi(x)$ found above, the current induced nonequilibrium electron concentration $n(x)$, see Eq.~(\ref{eq3}), behaves as $n(x)\propto 1/x$ for $|x|\gg l$, hence, the dipole moment of this charge distribution diverges. The {\it giant dipole moments} developing near the resistive elements of the circuit, and electrical fields around them, are an exceptional property of LD conductors. In 3D the screening is exponential and develops at the scale of $l_{3D}$. 

If a resistive conductor of the length $L$ and the conductivity $\sigma(x)={\rm const}$ is connected to two perfect conductors, then $\zeta(x<0)=0$, $\zeta(x)=jx/\sigma$ for $0<x<L$, and $\zeta(x)=jL/\sigma$ for $x>L$. Performing integrations in Eqs.~(\ref{eq5}) and (\ref{eq6}) results in two scales in $\varphi(x)$, $l$ and $L$. When $L\gg l$, the behavior of $\varphi(x)$ at the scale $L$ is controlled by the term $-(jl/\pi\sigma)\ln(|L-x|/|x|)$. It has the magnitude of about $(jl/\pi\sigma)\ln(L/l)$ and describes the electrical field $E_x(x)$ that penetrates into the perfect conductor (and the insulator around it) as deep as by several $L$. Fast changes in $\varphi(x)$ at the scale $l$ are of importance in the vicinity of the contacts, i.e., near $x=0$ and $x=L$. However, because the contacts are supposed be perfect, both $\varphi(x)$ and $E_x(x)$ are continuous near them.

In both model systems discussed above, the {\it Coulomb nonlocality} characteristic of LD conductors manifests itself quite distinctly. First, $j(x)$ and $E_x(x)$ are correlated at the scales of $l$ and $L$ that may be much larger than the scales $l_{\rm m.f.p.}$ and $l_{3D}$ typical of 3D conductors. Second, because of the logarithmic terms the problem of the potential distribution in a long homogeneous LD conductor has {\it no macroscopic limit}. It is only due to the 3D contacts fixing the potential drop that the total drops in $\varphi$ and $\zeta$ are equal, see below. Third, the electrical fields developing near the inhomogeneities of the conductor deeply penetrate the insulator surrounding it (or vacuum); they can be detected by a scanning microscope. These fields also result in the mechanical interaction of two LD conductors (their attraction or repulsion) and in their mechanical deformation. This mechanism provides a macroscopic approach to the current-induced forces investigated recently by first-principles calculations \cite{DVPL}.

The change in $E_x(x)$ in response to the modulation of $\sigma (x^\prime)$ in a remote point $x^\prime$ in the $j$= const regime, as well as the current induced distributions of $E_x(x)$ and $n(x)$ around the inhomogeneities, including the field in vacuum near the conductor, are the predictions of the theory that can be checked experimentally.

We are now in a position to investigate a realistic system of an inhomogeneous 2D conductor with a conductivity $\sigma(x)$ connected by perfect contacts to two classical conductors in the planes $x=0$ and $x=L$, their electrochemical potentials being $\zeta(0)=0$ and $\zeta(L)=\zeta_L$, respectively. Subtracting $x\zeta_L/L$ from $\zeta(x)$, we arrive at the function 
\begin{equation}
{\tilde \zeta}(x)=\zeta(x)-x\zeta_L/L 
\label{eq8}
\end{equation}
that equals zero at both ends of the interval $(0,L)$ and, therefore, can be expanded in a Fourier sine series 
  \begin{eqnarray}
{\tilde\zeta}(x)&=&\sum_{n=1}^\infty\left({2\over L}\int_0^L\zeta(u)\sin{{\pi nu}\over L}du+(-)^n{{2\zeta_L}\over {\pi n}}\right)\sin{{\pi nx}\over L}\nonumber\\
&=&{2\over\pi}\sum_{n=1}^\infty{1\over n}\sin{{\pi nx}\over L}
\int_0^L\zeta^{\prime}(u)\cos{{\pi nu}\over L}~du.
\label{eq9}
\end{eqnarray}
Here integration by parts was performed. Substituting Eq.~(\ref{eq9}) into Eq.~(\ref{eq5}) and (\ref{eq6}) results in the final expressions for $\psi(x,y)$ and $\varphi(x,y)$ in the upper half-strip:
\end{multicols}
\widetext
  \begin{equation}
\psi(x,y)={{\zeta_L}\over L}x+{2\over\pi}\sum_{n=1}^\infty{1\over n}\int_0^Ldu~ \zeta^{\prime}(u)\cos{{\pi nu}\over L}\sin{{\pi nx}\over L}\exp\left({-{{\pi ny}\over L}}\right),
\label{eq10}
\end{equation}
\begin{equation}
\varphi(x,y)=-{{\zeta_L}\over L}x-
{2\over\pi}\sum_{n=1}^\infty {1\over {n(1+\pi nl/L)}}
\int_0^Ldu~ \zeta^{\prime}(u)\cos{{\pi nu}\over L}\sin{{\pi nx}\over L}
\exp\left({-{{\pi ny}\over L}}\right).
\label{eq11}
\end{equation}
\begin{multicols}{2}
\noindent
The potential $\varphi(x,y)$ of Eq.~(\ref{eq11}) does not depend on $y$ in the planes $x=0$ and $x=L$, hence, the potential drop in the classical conductors\cite{RD02} being small compared to the potential drop in the LD conductor is disregarded. 

It follows from Eq.~(\ref{eq11}) that the typical penetration depth of the electrical field into the insulator is about $L$. However, $l$ controls the magnitude of the field that is reduced by the factor about $L/l$ when $\pi l/L\gg 1$.

Eq.~(\ref{eq11}) allows to find the kernel $K(x,u)=\delta E_x(x)/\delta\zeta^{\prime}(x)$ relating the response of the electrical field $E_x(x)=-\varphi^{\prime}(x)$ at the point $x$ to the variation of the resistivity $1/\sigma(u)=\zeta^{\prime}(u)/j$ at the point $u$. In 3D this relation is nearly local in the diffusion approximation, $K(x,u)\approx\delta(x-u)$ with the width of about $l_{3D}$. Differentiating Eq.~(\ref{eq11}) with respect to $x$, taking into account that $\zeta_L=\int_0^L\zeta^{\prime}(x)dx$, and performing the variational derivative, one finds 
\begin{equation}
K(x,y)={1\over L}+{2\over L}\sum_{n=1}^\infty{{\cos(\pi nx/L)\cos(\pi nu/L)}\over{1+\pi nl/L}}.
\label{eq12}
\end{equation}
This kernel is completely determined by the geometry of the system and does not depend on the specific form of $\sigma(x)$. The relation $\int_0^LK(x,u)~du=1$, providing for the equal drop of the electrical and electrochemical potentials across the specimen, follows from Eq.~(\ref{eq12}) because the integral from the second term vanishes. 

Using the identity $(1+\pi nl/L)^{-1}=\int_0^\infty d\lambda \exp[-\lambda(1+\pi nl/L)]$, the summation in Eq.~(\ref{eq12}) can be performed. Finally, $K(x,u)=1/L+f(x-u)+f(x+u)$ where
\begin{equation}
f(u)={1\over{2L}}\int_0^\infty d\lambda ~e^{-\lambda}{{\cos(\pi u/L)-\exp(-\pi l\lambda/L)}\over{\cosh(\pi l\lambda /L)-\cos(\pi u/L)}}.
\label{eq13}
\end{equation}
Integration in Eq.~(\ref{eq13}) over the region $\lambda\agt u/l$ results in a logarithmic singularity in $K(x,u)$ at $x\approx u$ that is much weaker than the $\delta(x-u)$ singularity in 3D. When the LD conductor is long enough, $l\alt L$, this singularity is of the scale $K(x,u)\sim (1/l)\ln(l/|x-u|)$, and is followed by the intermediate power-law asymptotic of $K(x,u)\sim 1/|x-u|$ in the region $l\ll |x-u|\ll L$ (if $l\ll L$). Under these conditions the overall scale of $K(x,u)$ is about $1/L$, however, for $|x-u|\alt l$ it reaches the scale $1/l$ and can exceed considerably the constant contribution $1/L$. In the opposite limit of a short LD conductor, $L\ll l$, the shape of $K(x,u)$ depends only on $L$ while its magnitude is proportional to $1/l$. Near the singularity $K(x,u)\sim (1/l)\ln(L/|x-u|)$. In this limit the screening comes mostly from the classical electrodes, and the term $1/L$ dominates the kernel $K(x,u)$.

Electrical field developing in the insulator around inhomogeneities of $\sigma(x)$ results in the electrostatic energy, and hence in the capacitance of LD conductors\cite{LL93}. We will show that for LD conductors $C$ has a giant magnitude and  shows a frequency dependence at relatively low frequencies $\omega$. For comparison, in 3D the capacitance can change at the scale of the geometric capacitance $C_g=\varepsilon/4\pi L$ (per unit area) when the modulation of $\sigma(x)$ is deep enough. However, a considerable frequency dependence of $C$ is expected only for frequencies larger than the inverse Maxwellian relaxation time, $\omega\agt\omega_{3D}=4\pi\sigma_{3D}/\varepsilon$, where $\sigma_{3d}$ is the 3D conductivity. With $\sigma_{3D}\approx\sigma/w$, one finds $\omega_{3D}\approx 4\pi \sigma/\varepsilon w$. Measuring the low-frequency dependence of the  capacity of a LD conductor might become a practical tool for detecting inhomogeneities of $\sigma(x)$.

For an external potential depending on the time $t$ as $\exp(-i\omega t)$, the continuity equation reads $j^{\prime}(x)+i\omega en(x)=0$. Eq.~(\ref{eq3}) with $\zeta(x)$ found from Eqs.~(\ref{eq8}) and (\ref{eq9}) and $\varphi(x)$ found from Eq.~(\ref{eq11}) should be used for $n(x)$. Finally, the continuity equation reads \cite{cond}
\begin{eqnarray}
&&d[\sigma(x)\zeta^{\prime}(x)]/dx \nonumber\\
&+&i{{\omega\varepsilon}\over{\pi L}}
\sum_{n=1}^\infty{{\sin(\pi nx/L)}\over{1+\pi nl/L}}
\int_0^L\zeta^{\prime}(u)\cos(\pi nu/L)~du =0.
\label{eq14}
\end{eqnarray} 
This is an integro-differential equation for $\zeta^{\prime}(x)$ that for $\omega\neq 0$ replaces the equation $\zeta^{\prime}(x)=j/\sigma(x)$ valid in the dc regime. The capacitance $C(\omega)$ and the active resistance $R(\omega)$, per unit width in the $z$ direction, that can be found from Eq.~(\ref{eq14}) depend on the frequency $\omega$. The characteristic frequency $\omega_{2D}\approx \sigma/\varepsilon l$ is controlled by the 2D screening length $l$, hence, $\omega_{2D}\ll \omega_{3D}$. The dispersion of $C(\omega)$ and $R(\omega)$ for $\omega\agt \omega_{2D}$ should allow to distinguish the capacity of a LD conductor from the geometrical capacitance.

The capacitance $C(\omega)$ is related to the total current in the circuit $J(t)$ including both $j(x,t)$ and the displacement current $j_d(x,y,t)$ originating from the charges accumulating at the electrodes. In a quasistationary regime the current $J(x,t)$ integrated over $y$ is independent of $x$
\begin{equation}
J(t)=\sigma(x)\partial\zeta(x,t)/\partial x+j_d(x,t)={\rm const}.
\label{eq15}
\end{equation}
The current $j_d(x,t)$ can be found from the second term of Eq.~(\ref{eq11}) that represents the contribution to $\varphi(x,y)$ coming from the area adjacent to the LD conductor:
\begin{eqnarray}
&&j_d(x)=(i\omega\varepsilon/4\pi)\int_{-\infty}^{\infty}dy~\partial\varphi(x,y)/\partial x\nonumber\\
&=&-{{i\omega\varepsilon}\over{\pi^2}}
\sum_{n=1}^\infty{{\cos(\pi nx/L)}\over{n(1+\pi nl/L)}}
\int_0^Ldu~\zeta^{\prime}(u)\cos({{\pi nu}\over L}).
\label{eq16}
\end{eqnarray}
In the lowest order in $\omega$, one can apply the static equation $\zeta^{\prime}(x)=\zeta_L/R_0\sigma(x)$ and express $j_d(x)$ in terms of the cosine-Fourier components of the resistivity $\sigma^{-1}(x)$
\begin{equation}
R_n=\int_0^Ldx~\cos(\pi nx/L)/\sigma(x),~~n\geq 0.
\label{eq17}
\end{equation}
Therefore
\begin{equation}
j_d(x)=-{{i\omega\varepsilon}\over{\pi^2}}{{\zeta_L}\over{R_0}}
\sum_{n=1}^\infty{{R_n\cos(\pi nx/L)}\over{n(1+\pi nl/L)}}~.
\label{eq18}
\end{equation}
Substituting Eq.~(\ref{eq18}) into Eq.~(\ref{eq15}), dividing it by $\sigma(x)$ and integrating over $x$ results in
\begin{equation}
JR_0=\zeta_L-{{i\omega\varepsilon}\over{\pi^2}}{{\zeta_L}\over {R_0}}
\sum_{n=1}^\infty {{R_n^2}\over{n(1+\pi nl/L)}}.
\label{eq19}
\end{equation}
Therefore, the final expression for the low-frequency capacitance (per unit width in the $z$ direction) is
\begin{equation}
C={\varepsilon\over{\pi^2}}\sum_{n=1}^\infty{{R_n^2/R_0^2}\over{n(1+\pi nl/L)}}~.
\label{eq20}
\end{equation}

Eq.~(\ref{eq20}) reveals a number of remarkable properties of the capacitance. First of all, $C$ depends on the conductivity $\sigma(x)$ only through the ratios $(R_n/R_0)^2$  [i.e., through the degree of the inhomogeneity of the LD conductor], is independent of the total resistance $R_0$, and vanishes for a homogeneous conductor\cite{note}. Second, $C\rightarrow 0$ when $l\rightarrow \infty$, hence, it comes exclusively from the screening in the LD conductor. Third, if $\sigma(x)$ is scaled as $\sigma(x/L)$, the ratios $R_n/R_0$ do not depend on $L$ while $C$ increases monotonically with $L$ and saturates as $L\rightarrow\infty$.

The most remarkable property of $C$ is its {\it giant magnitude} under the conditions when $l\alt L$ and the inhomogeneity is strong enough, $\sum_{n=1}^\infty R_n^2/n\sim R_0^2$. Indeed, in this case the ratio $C/C_g\sim L$. This implies that a 2D conductor with the thickness $w$ acquires an effective thickness $L\gg w$. This increase originates from the penetration of the electrical field into the insulator at the depth of about $L$. Only in the opposite limit, $l\gg L$, the ratio drops to $C/C_g\sim L^2/l$ because the field in the insulator is reduced by the factor $L/l$ as it is has already been mentioned above.

Giant capacitances are a general property of LD conductors that manifests itself quite dramatically for a resistive contact in a perfect conductor discussed above. With $\psi(x,y)$ of Eq.~(\ref{eq7}), the $x$ component of the electrical field in the $x=0$ plane equals $E_x(0,y)=-(j/\pi\Sigma l)\exp(y/l){\rm Ei}(-y/l)$, where ${\rm Ei}(-y/l)$ is the integral exponent function. For $y\gg l$ the field equals $E_x(0,y)\approx j/\pi y\Sigma $. Therefore, the integral over $y$ that enters into the displacement current $j_d(x=0)$, see Eq.~(\ref{eq16}), diverge logarithmically for large $y$, hence, the capacitance $C$ that is proportional to that integral, also diverges. This integral is cut-off by the geometry of the 3D conductors surrounding such a contact and screening it. Finally
\begin{equation}
C\approx (\varepsilon/2\pi^2)\ln(L/l)
\label{eq21}
\end{equation}
where $L$ is the cut-off length. 

Our approach to the Coulomb problem as developed for the diffusive transport is not restricted to it. In the classical regime, quantum wires should show properties similar to those of 2D systems but the mathematical techniques are different. As applied to the spin injection, the dimensional effects do not influence the two-terminal resistance controlled by $\zeta(x)$ but should be of importance for optics and transients because the Poisson equation is involved.

In conclusion, we have developed a technique for solving the Coulomb problem for 2D conductors and have investigated their properties including the strongly nonlocal relation between the resistivity and electric field, giant capacitances, and electrical fields developing around inhomogeneities. 

E.I.R. acknowledges the support from DARPA/SPINS by the Office of Naval Research Grant N000140010819.

\end{multicols} 

\begin{references}
\bibitem{note} This property is geometry dependent. A homogeneous conductor does not perturb the field produced by the electrodes (and, therefore, does not acquire capacitance) only when it is placed parallel to the force lines of that field. This observation suggests that in a more general geometry (curved electrodes or a curved LD conductor) even a homogeneous LD conductor acquires capacitance.
\end{references}
\end{document}